\documentclass[preprint,aps,showkeys,groupedaddress,longbibliography,prd,floatfix,11pt]{revtex4-1}

\usepackage{amsfonts}
\usepackage{amssymb}
\usepackage{amsmath}
\usepackage{xcolor}
\usepackage{soul} 
\usepackage{ulem}
\usepackage{enumitem}

\usepackage{lipsum}
\usepackage{mathptmx}
\usepackage{mathtools}
\usepackage{lmodern}

\providecommand{\di}{{}^{\mbox{\tiny$(n)$}}\!}

\begin{document}

\title{Kaluza-Klein Dimensional Reduction From Elasticity Theory of Crumpled Paper}
%\title{Kaluza-Klein Theory of Gravitation and Electromagnetism from Elasticity Theory of Crumpled Paper}

\author{Mokhtar Adda-Bedia}\email{mokhtar.adda-bedia@ens-lyon.fr}
\affiliation{Universit\'e de Lyon, Ecole Normale Sup\'erieure de Lyon, CNRS, Laboratoire de Physique, F-69342 Lyon, France}
\author{Eytan Katzav}\email{eytan.katzav@mail.huji.ac.il}
\affiliation{The Racah Institute of Physics, The Hebrew University of Jerusalem, Jerusalem 9190401, Israel}

\date{\today}

\begin{abstract}

During the last century, two independent theories using the concept of dimensional reduction have been developed independently. The first, known as F\"oppl-von K\'arm\'an theory, uses Riemannian geometry and continuum mechanics to study the shaping of thin elastic structures which could become as complex as crumpled paper. The second one, known as Kaluza-Klein theory, uses Minkowskian geometry and general relativity to unify fundamental interactions and gravity under the same formalism. Here we draw a parallel between these two theories in an attempt to use concepts from elasticity theory of plates to recover the Einstein-Maxwell equations. We argue that Kaluza-Klein theory belongs to the same conceptual group of theories as three-dimensional elasticity, which upon dimensional reduction leads to the F\"oppl-von K\'arm\'an theory of two-dimensional elastic plates. We exploit this analogy to develop an alternative Kaluza-Klein formalism in the framework of elasticity theory in which the gravitational and electromagnetic fields are respectively associated with stretching-like and bending-like deformations. We show that our approach of dimensional reduction allows us to retrieve the Lagrangian densities of both gravitational, electromagnetic and Dirac spinors fields as well as the Lagrangian densities of mass and charge sources. 

\end{abstract}

%\keywords{}

\maketitle

\section{Introduction}

In their seminal work, Kaluza~\cite{Kaluza1921} and Klein~\cite{Klein1926} provided a scheme to unify electromagnetism and gravity by supplementing general relativity with an additional fifth dimension of spacetime. The ideas of cylindrical boundary condition in Kaluza and compactification in Klein state that the extra dimension, involving the electromagnetic field, would be physically undetectable even though it would provide a means of unification. Since then, the Kaluza-Klein (KK) paradigm is often at the basis of physical theories that aim to unify the known fundamental interactions (electromagnetic, weak nuclear and strong nuclear) with gravity as well as to generate elementary particles from common first principles. Such a quest has seen a revival with the emergence of superstring and supergravity theories \cite{Cho1975,Witten1981,Mecklenburg1984,Duff1986,Bailin1987,Overduin1997,Wesson,Wesson2006,Zee2013}, which are based on the idea that the fundamental building blocks of nature are vibrating modes of strings and membranes in high dimensions.

Dimensional reduction is not unique to KK-like theories. In mechanics of continuous media, a similar procedure is performed to model geometrical and mechanical response of elastic membranes from elasticity equations of three-dimensional bulk materials. While the general geometrical foundations were developed by Gauss \cite{Millman1977}, the formal basis for understanding strongly deformed plates came from the work of F\"oppl~\cite{foppl1921vorlesungen} and von K\'arm\'an~\cite{von1956collected}. The constraints of mechanical equilibrium and differential geometry were exploited to reduce the degrees of freedom of a sheet onto two scalar fields associated with the curvature tensor and the two-dimensional stress tensor, respectively. The former mediates the out-of-plane bending deformations and the latter the in-plane stretching deformations. The nonlinear coupling between the two fields in the resulting F\"oppl-von K\'arm\'an (FvK) equations often leads to localized buckling events which are precursors to patterns similar to those formed in crumpled paper \cite{BenAmar1997,lobkovsky1997properties}. For the last three decades, various aspects related to FvK equations  and their extensions have been the area of intense research from both mechanical, physical and mathematical perspective \cite{Kramer1997b,Witten2007,Efrati2009,Audoly}.

Although the underlying geometries are different, Riemannian geometry for elastic plates and Minkowski geometry for general relativity, we seek here to relate these two areas in which the concept of dimensional reduction was developed. First, we identify a scale separation between gravitational and electromagnetic modes of the Einstein-Maxwell (EM) action, which is similar to that between stretching and bending energy densities involved in describing the elastic response of thin plates. Consequently, we argue that KK formalism is of the same kind as the reduction of three-dimensional (3D) elasticity equations leading to the FvK equations of 2D elastic plates. Then, we exploit this analogy to develop a KK formalism in the framework of elasticity theory of thin plates in which the gravitational and electromagnetic fields are respectively associated with stretching-like and bending-like deformations.

This paper is organized as follow. We start by drawing the similarity between FvK and EM formalisms which allows us to propose an elastic-like approach to KK theory. To remain self-consistent we relegate the introduction of basic elements of elasticity theory of thin plates to Appendix~\ref{app:FvK}.  Using this analogy, we perform  a dimensional reduction of the action of 5D gravitational and matter fields which we consider as different entities. The resulting 4D action involves a single additional term which reflects an explicit interaction between gravitational and electromagnetic fields and which contributes only at scales within the matter content. To verify the validity of our approach at quantum scales, we also apply the same scheme of dimensional reduction to the action of Dirac spinor field. Finally, we conclude by pointing out the main results and possible extensions of our approach.

To keep it concise, most of the algebraic computations are relegated to Appendices. Moreover, even if it is not adequate {\it stricto sensu}, we will sometimes use a language of mechanics in general relativity to reinforce the analogy elasticity-general relativity. Finally, throughout the whole article including the appendices, we shall adopt notational conventions to distinguish the quantities defined in 5D from their 4D equivalent. Unless otherwise specified, functions defined in five dimensions are noted by the same letters used for four ordinary space time with a circumflex accent. We employ the alphabetic convention that letters $M,N,..$ (resp. $\mu,\nu,...$) will denote world indices taking the values $0,1,2,3,4$ (resp. $0,1,2,3$). Finally, when using the tetrad representation the indices from the early alphabet $\mathrm{A},\mathrm{B},...$ (resp. $\mathrm{a},\mathrm{b},...$) shall be the frame labels taking the values $0,1,2,3,4$ (resp. $0,1,2,3$).

\section{Similarity of F\"oppl-von K\'arm\'an and Einstein-Maxwell formalisms}

In 4D spacetime, the actions that corresponds to Einstein gravity and Maxwell electromagnetism are given by~\cite{LL1980}
\begin{eqnarray}
{{S}_\mathrm{E}}  &=& -\frac{c^3}{16\pi G}\int \mathrm{d}V\sqrt { - g} R\;,
\label{eq:SEinstein} \\
{{S}_\mathrm{M}} &=& -\frac{1}{16\pi c}\int \mathrm{d}V\sqrt { - g}F^{\mu \nu }F_{\mu \nu }\;,
\label{eq:SMaxwell}
\end{eqnarray}
where $c$ is the speed of light, $G$ is the gravitational constant, $\mathrm{d}V$ is the 4D volume element in hyperspace, $R$ is the Ricci (scalar) curvature of the Lorentzian metric $g_{\mu\nu}$, and $F_{\mu\nu}$ is the electromagnetic field tensor given by $F_{\mu\nu} =A_{\nu;\mu}-A_{\mu;\nu}$ where $A_\mu$ denotes the vector potential field. While the two theories have been developed separately, the classical Einstein-Maxwell action ${{S}_\mathrm{EM}}$ is simply defined as the sum of the two ${S}_\mathrm{E} +{S}_\mathrm{M}$, namely~\cite{LL1980}
\begin{equation}
{{S}_\mathrm{EM}} = -\frac{c^3}{16\pi G}\int \mathrm{d}V\sqrt { - g}\left(   R +\frac{G}{ c^4}  F^{\mu \nu }F_{\mu \nu }\right)\;.
\label{eq:S0_fM}
\end{equation}

KK type theories aim to recover the action given by Eq.~(\ref{eq:S0_fM}) from an underlying unified theory involving both gravitational and electromagnetic potential fields. Similarly, the total action of mass and electromagnetic current is the sum of the actions of two source terms~\cite{LL1980}
\begin{equation}
{{S}_{m}} = \int\limits_{}^{} \mathrm{d}V\sqrt { - g} \rho_m c - \frac{1}{c} \int\limits_{}^{}\mathrm{d}V\sqrt { - g} \rho_e u^\mu A_\mu\;,
\label{eq:12ter}
\end{equation}
where $u^\mu$ is the velocity vector along the worldline of the matter, $\rho_m$ and $\rho_e$ are mass and charge density distributions per unit space three-volume. Any definitive unifying theory should also justify the additivity of the two source terms leading to Eq.~(\ref{eq:12ter}). Such requisite is usually a discriminating condition because it is more difficult to achieve. Here we argue that we can achieve such a goal using an approach borrowed from elasticity theory of continuous media. In particular, we state that the KK compactification procedure is reminiscent of the dimensional reduction of elasticity equations from three-dimensional (3D) bulk materials to two-dimensional (2D) plates yielding FvK equations. For reasons of self-consistency, the details of the latter theory are summarized in Appendix~\ref{app:FvK}.

Let us start by deriving scaling properties associated with the different terms involved in Eqs.~(\ref{eq:S0_fM})-(\ref{eq:12ter}). Without loss of generality, we assume that both mass and charge densities follow the same spatial distribution, which is a realistic assumption for localised matter content~\cite{LL1980}. Therefore, we can define a mass scale $m$ and a charge scale $q$ associated to constituents of matter such that $\rho_e/\rho_m=q/m$. This allows us to define characteristic length scales given by
\begin{equation}
\mathrm{r}_m=\frac{Gm}{c^2}\;,
\quad \mathrm{r}_e=\frac{\rho_e q}{\rho_mc^2}=\frac{q^2}{mc^2}\;,
\quad \sqrt{\mathrm{r}_m\mathrm{r}_e}=\sqrt{\frac{Gq^2}{c^4}}\;,
\label{eq:lengths}
\end{equation}
where $\mathrm{r}_m$ and $\mathrm{r}_e$ are length scales associated with the content of matter and its nature (mass and charge). In addition, we can define the geometric length scale $L$ associated with the range over which the 4D gravitational and potential fields vary. $L$ can be seen as the characteristic extension of the 4D hyperspace, and should satisfy $L\gg \{ \mathrm{r}_m,\mathrm{r}_e \}$ as long as matter consists of localized sources. As a first approximation, one can assume that variations of the metric $g$ and the potential $A_\mu$ are induced independently by the mass $m$ and the charge $q$ respectively. In this case, dimensional analysis allows us to deduce that these two fields obey scaling properties such that $g\sim \mathrm{r}_m/L$ and $A_\mu\sim q/L$. Consequently, the Ricci scalar and the field tensor scale as $R\sim  \mathrm{r}_m/L^3$ and $F_{\mu\nu}\sim  q/L^2$. Therefore, using Eq.~(\ref{eq:lengths}) one shows that the two components of the action in Eq.~(\ref{eq:S0_fM}) satisfy 
\begin{equation}
\frac{G}{c^{4}}F^{\mu\nu}F_{\mu\nu}\sim \left(\frac{\mathrm{r}_e}{L}\right) R\ll R \;.
\label{eq:scale-sep}
\end{equation}
Eq.~(\ref{eq:scale-sep}) shows that there is a {\it scale separation} between gravitational and electromagnetic effects as long as $\mathrm{r}_e\ll L$.

At this stage we draw an analogy with elasticity theory of plates. The action ${S}_\mathrm{EM}$ given by Eq.~(\ref{eq:S0_fM}) has the same structure as the total elastic energy of a plate of thickness $t$ which involves two distinct contributions quantifying its stretching and bending energies. Whereas in elastic plates the ratio of stretching over bending energies is quadratic in $(t/L)\ll1$ (see Appendix~\ref{app:FvK}), here the gravitational over electromagnetic Lagrangian densities scale linearly in $(\mathrm{r}_e/L)\ll1$. Except for this difference in the strength of scale separation, KK dimensional reduction of 5D gravity to 4D gravity and electromagnetism recalls FvK dimensional reduction of 3D elasticity equations. In view of this analogy, the gravitational term $R$ can be interpreted as a stretching-like response of a 5D gravity field within a 4D hypersurface and the electromagnetic term $\frac{G}{c^{4}}F^{\mu\nu}F_{\mu\nu}$ as bending-like response. In other words, the 4D metric field $g_{\mu\nu}$ describes internal 4D deformations while the potential field $A_\mu$ encodes deformations along the fifth dimension.

In the following, we lay the groundwork for our approach using the analogy described above. It consists of assuming that the potential field $A_\mu$ arises as a {\it perturbation} through the fifth dimension around a 4D ground state that involves only the 4D gravitational field. Specifically, we perform an expansion of a 5D metric whose leading order yields pure gravitational Einstein equations and the electromagnetic field show up as a first order perturbation whose magnitude is proportional to the ``thickness'' of the fifth direction of spacetime. Notice that although our approach resembles KK dimensional reduction, it is different from classical KK theory in that it does not assume neither compactification nor  cylindrical boundary condition.

\section{An elastic-like approach to Kaluza-Klein theory}

Following~\cite{MAB}, we consider a general multidimensional pseudo-Riemannian metric built upon a 5D Lorentzian spacetime metric $\hat{g}$ and an infinite number of Euclidean-like diagonal components such that the distance element along a worldline $\mathrm{d}s$ is given by
\begin{equation}
\mathrm{d}s^2=\hat{g}_{MN} \mathrm{d}x^M \mathrm{d}x^N- \sum_{i=5}^\infty \mathrm{d}z^i\mathrm{d}z^i\;.
\label{eq:metric0}
\end{equation}
Here, $\hat{g}_{MN}$ depends on the coordinates $x^M$ only. We refer to $x^M$ as the active coordinates and $z^i$ ($i\geq 5$) as the passive ones. We define the action $\hat{S}_G$ of the gravitational field associated with the metric $\hat{g}$ as a generalization of the Einstein-Hilbert action to a 5D spacetime given by~\cite{Overduin1997,MAB}
\begin{equation}
\hat{S}_\mathrm{G}=  - \frac{{{c^3\ell^{-1}}}}{{16\pi G}}\int\limits_{}^{} {\mathrm{d}\hat{V}\sqrt{\left|\hat{g}\right|}  \hat R} \;,
\label{eq:action0}
\end{equation}
where $\hat{R}$ is the Ricci curvature of the metric given by Eq.~(\ref{eq:metric0}), $\mathrm{d}\hat{V}$ is the volume element of the 5D hyperspace associated with the metric $\hat{g}$, and $\ell$ is a length scale needed to render the action carry the physical dimensions $[M][L]^2[T]^{-1}$.

Using the analogy with dimensional reduction of 3D elasticity theory to 2D thin plate elasticity, yielding the FvK equations, we aim at developing a KK-like formalism that yields Einstein-Maxwell equations starting from a 5D description of gravity. For this, we assume that the ground state results from  a purely ``stretching-induced'' gravitational field described by a 4D active spacetime. Then, we perform ``out-of-plane'' perturbations around this zeroth order state that are induced by $x^4$-dependent fluctuations of the gravitational field. 

The leading order of such an expansion should recover the classical 4D Einstein-Hilbert action given by Eq.~(\ref{eq:SEinstein}), with $\zeta\equiv x^4$ behaving as a passive coordinate. The latter condition imposes that the zeroth order components of the metric $\hat{g}$ should satisfy $\hat{g}_{M4}=-\delta_{M4}$ and $g_{\mu\nu}$ should be independent of $\zeta$. Such a metric yields $\hat R=R$ and $\left|\hat{g}\right|=\left|g\right|$ and therefore, to retrieve $\hat S_G =S_E$ from Eq.~(\ref{eq:action0}) one should impose $\int d\zeta =\ell$. This result holds for any partition of the hyperspace between active and passive components~\cite{MAB}. The Einstein-Hilbert action generalized to any spacetime dimension introduces a single physical constant $\ell$ characterizing the extension of the hyperspace in each passive direction of the metric of Eq. (\ref{eq:metric0}). The magnitude of $\ell$ is unknown since scaling arguments do not invoke any quantum property that could allow us to identify $\ell$ with the Planck length $\ell_P$.

Again, using the analogy with the theory of elasticity of slender structures, the ground state should be translationally invariant in the direction of the fifth spatial dimension. It is described by a 4D spacetime metric $g$ that lies on a centroid: a hypersurface defined by a worldline $\zeta=0$. Due to the finite ``thickness" $\ell$ of the hyperspace, perturbations around this ground state are induced by variations of a 5D metric such that $\hat{g}(x^M)=g(x^\mu) +\sum_n \di g(x^\mu)\zeta^n$, with $|\zeta|<\ell /2$. That is, we assume that the perturbations of the metric across the thickness do not modify the ``geometrical'' structure of the hyperspace.  This allows us to expand to any desired order in $\zeta$ the gravitational action~(\ref{eq:action0}) which can be rewritten as
\begin{equation}
\hat{S}_\mathrm{G}= - \frac{{{c^3\ell^{-1}}}}{{16\pi G}} \int\limits_{ - \ell/2}^{\ell/2} {\mathrm{d}\zeta\int\limits_{}^{} {\mathrm{d}V\sqrt{\left|\hat{g}\right|}  \hat R}} \;.
\label{eq:action1}
\end{equation}
In the following, we start from a general excpansion of $\hat g$ in powers of $\zeta$ and compute the corresponding expansion of the action $\hat{S}_\mathrm{G}$. Next, we identify the specific expansion of the 5D metric that allows us to get as close as possible to the usual EM action $S_\mathrm{EM}$ given by Eq.~(\ref{eq:S0_fM}) from the gravitational action $\hat S_\mathrm{G}$ given by Eq.~(\ref{eq:action1}). Next, we apply the resulting metric to 5D matter and Dirac spinor fields to determine the conditions under which their corresponding 4D definitions could be retrieved.

\subsection{Dimensional reduction of the gravitational field}

We aim at determining the most general metric ${\hat{g} _{MN}}$ such that the integrand of the action $\hat S_\mathrm{G}$ in Eq.~(\ref{eq:action1}) would be of second order in $\zeta$. To this purpose we start with the expansion of the metric
\begin{equation}
{\hat{g} _{MN}} = \left( {\begin{array}{*{20}{c}}
{{g_{\mu \nu }} + a{\zeta^2}{A_\mu }{A_\nu } + \bar a{\zeta^4}{{A^2}} {A_\mu }{A_\nu }}&{b\zeta{A_\mu } + \bar b{\zeta^3}{{A^2}} {A_\mu }}\\
{b\zeta{A_\mu } + \bar b{\zeta^3} {{A^2}} {A_\mu }}&{ - 1 + c{\zeta^2} {{A^2}} + \bar c{\zeta^4}{{{A^4}}}} 
\end{array}} \right)+{\cal O}\left(\zeta^5\right)\;.
\label{eq:1}
\end{equation}
where $g_{\mu\nu}$ is the 4D metric tensor, $A_\mu$ is an electromagnetic-like potential field ($A^2=A_\mu A^\mu$) and $(a,b,c,\bar{a},\bar{b},\bar{c})$ are physical constants. Notice that the constants used in this Section and Appendix~\ref{Perturb_KK} should not be confused with other physical constants and indices. Eq.~(\ref{eq:1}) is the most general expansion of the 5D metric that involves a single vector field and whose leading order reduces to a 4D metric with a fifth-dimensional passive component. Notice that the metric elements are dimensionless and the physical dimensions of $A_\mu$ are $[e][L]^{-1}$. Thus, considering only powers of the combination $\zeta A_\mu$ allows us to avoid introducing characteristic length scales in the metric elements (the charge scale can be absorbed in the multiplicative constants), apart from the geometrical scale $\ell$ implied by the passive components of hyperspace.
 
In Appendix~\ref{Perturb_KK} we perform expansions in powers of $\zeta$ of the determinant of $\hat g _{MN}$, the 5D Christoffel symbols $\hat \Gamma _{MN}^R$, the curvature tensor $\hat R _{MN}$ and the Ricci scalar $\hat R$. We find that the first correction to the usual $\sqrt{-g} R$ in the integral term of the action $\hat S_\mathrm{G}$ is of order $\zeta^2$. A careful analysis shows that if one requires that under the integral sign of the action there must stand an expression quadratic in the field $A_\mu$ which involves only the derivatives $A_{\mu;\nu}$, then one should impose $a = \bar a =  0$, as well as $c = {b^2}$. It also turns out that neither $\bar b$ nor $\bar c$ contribute to the second order in $\zeta$ in the integrand of the action. Hence, in the following we restrict our discussion to the simpler 5D metric 
\begin{equation}
{\hat{g} _{MN}} = \left( {\begin{array}{*{20}{c}}
{{g_{\mu \nu }}}&{b\zeta{A_\mu }}\\
{b\zeta{A_\mu }}&{ - 1 + {b^2}{\zeta^2}{A^2}}
\end{array}} \right)
+{\cal O}\left(\zeta^5\right) \;,
\label{eq:2}
\end{equation}
which leaves a single unknown parameter $b$. The determinant of this metric obeys (see Appendix~\ref{Perturb_KK})
\begin{equation}
\sqrt{\left|\hat{g}\right|}  = \sqrt { - g} +{\cal O}\left(\zeta^3 \right)\;,
\label{eq:10bis}
\end{equation}
while its Ricci scalar is
\begin{equation}
\hat R =  R + \tfrac{3}{4}{b^2}{\zeta^2}{F^{\mu \nu }}{F_{\mu \nu }} - {b^2}{\zeta^2}{R_{\mu \nu }}{A^\mu }{A^\nu }+ {\cal O}\left(\zeta^3 \right)\;.
\label{eq:10}
\end{equation}
Notice that the total derivatives in the expression for $\hat R$ have been discarded assuming that boundary terms do not contribute to the action. After integration over $\zeta$, Eq.~(\ref{eq:action1}) transforms into
\begin{equation}
\hat{S}_\mathrm{G} =- \frac{{{c^3}}}{{16\pi G}}\int\limits_{}^{} \mathrm{d}V \sqrt { - g}R- \frac{b^2\ell^2 c^3}{{256\pi G}}\int\limits_{}^{} {\mathrm{d}V\sqrt { - g} \left( {{F^{\mu \nu }}{F_{\mu \nu }} - \frac{4}{3}{R_{\mu \nu }}{A^\mu }{A^\nu }} \right)} + {\cal O}\left(\ell^3\right)\;.
\label{eq:11}
\end{equation}
The sign of the coefficient in front of $F^{\mu \nu }F_{\mu \nu }$ is negative definite just like for the Maxwell action $S_\mathrm{M}$ in Eq.~(\ref{eq:SMaxwell}). 
Imposing an equality between the coefficients of $F^{\mu \nu }F_{\mu \nu }$ in Eq.~(\ref{eq:SMaxwell}) and Eq.~(\ref{eq:11}) yields 
\begin{equation}
b\ell= \frac{4 \sqrt{G}}{c^2}=\frac{4\sqrt \alpha  }{e} \,{\ell _p}\;,
\label{eq:bl}
\end{equation}
where $\alpha=e^2/\hbar c$ is the fine structure constant and ${{\ell }_{p}}=\sqrt{{\hbar G}/{{{c}^{3}}}\;}$ is the Planck length. With this result, all the coefficients introduced in the expansion~(\ref{eq:1}) of the metric are determined. Eq.~(\ref{eq:bl}) shows that $b\ell\propto\ell_P$, which is not sufficient to identify $\ell$ with $\ell_P$. However, it shows that $|b\zeta A_\mu|<b\ell \sqrt{A^2} \sim\ell_p/L\ll1$, where $L$ is the length scale associated with the variation of the vector field $A_\mu$. This result justifies a posteriori our perturbative approach and shows that it is not constrained by the hypothesis of a slender
fifth dimension since the ratio $\ell/\ell_P$ is not constrained.

Our approach produces an additional term in the action when compared with Eq.~(\ref{eq:S0_fM}) which is proportional to ${R_{\mu \nu }}{A^\mu }{A^\nu }$ and breaks the gauge invariance that is usually associated with the vector potential $A_\mu$. The existence of this term is justified thanks to the identity~(\ref{eq:A27}), which allows to rewrite it as a sum of terms that involve covariant derivatives of $A_{\mu;\nu}$ and total derivatives. This explicit gravitational-electromagnetic interaction term does not appear neither in Einstein-Maxwell action~(\ref{eq:S0_fM}) nor in the classical KK theory. However, it obeys the weak principle of equivalence and thus does not violate the gauge invariance at the level of special relativity~\cite{Li2016}.

Before discussing further the results of this section, we consider the perturbation of a 5D gravitational source field within a metric given by Eq.~(\ref{eq:2}).

\subsection{Dimensional reduction of the matter field}
\label{sec:matter}

We intend to generalize the elastic-like approach to KK theory in order to model gravity and electromagnetism in the presence of matter fields. To this aim, we need to specify the nature of the 5D energy density of source terms. We assume that the corresponding action is associated with a matter component only. Let us then start with an action $\hat S_m$ given by
\begin{equation}
{\hat{S}_{m}} = \int\limits_{ - \ell/2}^{\ell/2} \mathrm{d}\zeta\int\limits_{}^{} \mathrm{d}V\sqrt { |\hat g|} \hat \rho c \;,
\label{eq:12}
\end{equation}
where $\hat \rho$ is the mass density of the body, i.e. mass per unit space four-volume. We emphasize that the absolute mass density should be defined per unit proper four-volume, that is the volume in the reference system in which the given portion of the body is at rest~\cite{LL1980}. More precisely, one has
\begin{equation}
\hat \rho=\frac{\hat\varrho}{\sqrt{\hat g _{00}}}\frac{\mathrm{d}\hat s}{\mathrm{d}x^0}\;,
\end{equation}
where $\hat \varrho$ is the absolute mass density per unit space four-volume 
\cite{LL1980} and $\mathrm{d}\hat s$ is the infinitesimal line element along the 5D worldline of the matter. We assume that mass is distributed across the $\zeta$-direction such that $\hat \varrho(x^M) =\kappa(\zeta)\varrho(x^\mu)$. Using $\hat g _{00}\approx g _{00}$, which is correct to first order in $\zeta$ (see Eq.~\ref{eq:2}), one gets 
\begin{equation}
\hat \rho(x^M)= \kappa(\zeta)\,\rho_m(x^\mu)\frac{\mathrm{d}\hat s}{\mathrm{d} s}\;,
\label{eq:rho5}
\end{equation}
where $\kappa(\zeta)$ is an unknown function of physical dimension $[L]^{-1}$ that describes the mass distribution in the fifth dimension, $\rho_m$ is the mass density distribution per unit space three-volume and $\mathrm{d} s$ is the infinitesimal line element along the 4D worldline of the matter~\cite{LL1980}.

Eq.~(\ref{eq:rho5}) shows that $\hat \rho(x^M)$ depends explicitly on the field $A_\mu$ through the term $\mathrm{d}\hat s/\mathrm{d} s$ which is given by
\begin{equation}
 \frac{\mathrm{d}\hat s}{\mathrm{d} s}=\sqrt{\frac{\hat g_{MN}\mathrm{d}x^M\mathrm{d}x^M}{\mathrm{d}s^2}}=\sqrt{1+2b\zeta u^5 u^\mu A_\mu -\left(1-b^2\zeta^2 A^2\right)u^5u^5}\;,
 \label{eq:worldline}
\end{equation}
where $u^\mu=\frac{\mathrm{d}x^\mu}{\mathrm{d}s}$ and $u^5=\frac{\mathrm{d}\zeta}{\mathrm{d}s}$ are the components of the $5$-velocity field. Let us assume a priori that $u^5$ depends on the component $\zeta$ only and satisfies $u^5= -\beta \zeta+O(\zeta^3)$ with $0<\beta\ell\ll 1$. Expanding Eq.~(\ref{eq:worldline}) up  to second order in $\zeta$ and substituting in Eq.~(\ref{eq:12}) yields
\begin{equation}
{\hat{S}_{m}} = \int\limits_{}^{} \mathrm{d}V\sqrt { - g} \rho_m c \int\limits_{ - \ell/2}^{\ell/2} \mathrm{d}\zeta \kappa(\zeta)\left[1- \frac{1}{2}(2b\beta u^\mu A_\mu+\beta^2) \zeta^2 + {\cal O} \left(\zeta^3\right)\right]\;.
\label{eq:12bis}
\end{equation}
One should compare Eq.~(\ref{eq:12bis}) with the usual 4D source terms given by Eq.~(\ref{eq:12ter}). It is straightforward to show that Eqs.~(\ref{eq:12ter}) and (\ref{eq:12bis}) are equivalent if $\kappa(\zeta)$ satisfies the identities
\begin{eqnarray}
\int\limits_{ - \ell/2}^{\ell/2}  \kappa(\zeta) \mathrm{d}\zeta &=&1+\frac{\beta \rho_e}{2 b \rho_m c^2}\;,
\label{eq:kappa1}\\
\int\limits_{ - \ell/2}^{\ell/2}   \zeta^2\kappa(\zeta)\mathrm{d}\zeta &=&\frac{\rho_e}{b\beta \rho_m c^2}\;.
\label{eq:kappa2}
\end{eqnarray}
Eqs.~(\ref{eq:kappa1})-(\ref{eq:kappa2}) are physically relevant if both the mass $\rho_m(x^\mu)$ and the charge $\rho_e(x^\mu)$ densities follow the same distribution. This is a realistic property which particularly allows us to define a characteristic length scale $\mathrm{r}_e$ associated with the source fields as proposed by Eq.~(\ref{eq:lengths}). A constant uniform $\kappa(\zeta)$ satisfies Eqs.~(\ref{eq:kappa1},\ref{eq:kappa2}), however such a solution is consistent only for $\beta\ell$ of order unity, which contradicts the initial assumption $\beta\ell\ll1$. Nevertheless, a nonuniform $\kappa(\zeta)$ could satisfy Eqs.~(\ref{eq:kappa1})-(\ref{eq:kappa2}) without selecting the parameter $\beta \ell$. An example for such a solution is given by
\begin{equation}
\kappa(\zeta)=  \left(1+\frac{\beta \rho_e}{2 b \rho_m c^2}\right)\delta(\zeta)+\frac{2 \rho_e}{b\beta \ell^2 \rho_m c^2} 
\left[\delta \left(\zeta-\ell/2 \right)-2\delta(\zeta)+\delta \left(\zeta+\ell/2 \right) \right] \;.
\label{eq:kappa3}
\end{equation}
Eq.~(\ref{eq:kappa3}) shows that $\kappa(\zeta)$ depends explicitly on $\beta \ell$ which is still an unknown dimensionless quantity although we expect $\beta \ell\propto b\ll 1$. Our approach only provides a relation between the two independent quantities $\kappa(\zeta)$ and $u^5(\zeta)$ without indicating a principle to select them independently. Despite this arbitrariness, we provide a unifying scenario for the origin of two manifestations of real-world matter (mass and charge): they arise from a single matter-like energy density distribution in the 5D spacetime which manifests itself in the 4D spacetime between mass and charge. In the example of Eq.~(\ref{eq:kappa3}), the first term results from a monopole and the latter terms from a quadrupolar distribution.

\subsection{Gravity and electromagnetic field equations}
 
Combining the results of the dimensional reduction of 5D gravity and matter fields, we arrive at a total action $S_{\mathrm{T}}={\hat{S}_{\mathrm{G}}}+{\hat{S}_{m}}$ given by
\begin{eqnarray}
S_{\mathrm{T}} = &&- \frac{{{c^3}}}{{16\pi G}}\int\limits_{}^{} \mathrm{d}V\sqrt { - g}R+\int\limits_{}^{} \mathrm{d}V\sqrt { - g} \rho_m c\nonumber\\
&&-\frac{1}{16\pi c}\int\limits_{}^{} {\mathrm{d}V\sqrt { - g} \left( {{F^{\mu \nu }}{F_{\mu \nu }} - \frac{4}{3}{R_{\mu \nu }}{A^\mu }{A^\nu }} \right)}
 - \frac{1}{c^2} \int\limits_{}^{} \mathrm{d}V\sqrt { - g} J^\mu A_\mu\;,
\label{eq:S_fM}
\end{eqnarray}
where $J^\mu=\rho_e c u^\mu$ is the 4D current vector. Eq.~(\ref{eq:S_fM}) recovers the classical EM action in addition to a single term proportional to ${R_{\mu \nu }}{A^\mu }{A^\nu }$ which suggests an explicit interaction between gravity and EM fields. Using similar scaling arguments as the ones leading to Eq.~(\ref{eq:scale-sep}), we can show that
\begin{equation}
{R_{\mu \nu }}{A^\mu }{A^\nu }\sim  \left(\frac{\mathrm{r}_m}{L}\right) F^{\mu\nu}F_{\mu\nu}\ll F^{\mu\nu}F_{\mu\nu}\;.
\label{eq:scale-sep2}
\end{equation}
Therefore, the new interaction term is always negligible compared to the strength of both the electromagnetic Lagrangian density and the interaction term $J^\mu A_\mu/c$.

To minimize the total action given by Eq.~(\ref{eq:S_fM}), we use again the analogy with elastic plates. Without implying any geometric meaning, the fields $g_{\mu\nu}$ and $A_\mu$ can be viewed as in-plane and out-of-plane fields of a 4D manifold. However, in contrast to elastic plates which are Euclidean surfaces constrained by Gauss's {\it Theorema Egregium} (see Appendix A), these two fields are independent variables since they operate within a Lorentzian manifold. Therefore, the minimization of the action $S_{\mathrm{T}}$ should be performed with respect to these two variables without additional constraints. Since the electromagnetic terms depend on the metric, a full minimization of the action with respect to $g_{\mu\nu}$ yields cumbersome modified Einstein equations~\cite{Li2016}. Nevertheless, the scale separation between gravitational and electromagnetic contributions highlighted by Eqs.~(\ref{eq:scale-sep}),(\ref{eq:scale-sep2}) allows us to settle for a perturbative scheme. First, we minimize the stretching terms of $S_{\mathrm{T}}$ (the ones that do not involve $A_\mu$) with respect to $g_{\mu\nu}$ and then the bending terms with respect to $A_\mu$. This procedure gives the following two equations
\begin{eqnarray}
&&R^{\mu\nu}-\frac{1}{2}g^{\mu\nu}R= \frac{8\pi G}{c^3}T^{\mu\nu}\;,
\label{eq:Efinal}\\
&&F^{\mu\nu}_{\;\;\;\;;\nu}-\frac{2}{3} R^{\mu\nu}A_\nu=-\frac{4\pi}{c} J^\mu \;,
\label{eq:Mfinal}
\end{eqnarray}
where $T^{\mu\nu}=\rho_m c u^\mu u^\nu$ is the energy momentum tensor of pure matter content~\cite{LL1980}. Eq.~(\ref{eq:Efinal}) is the usual Einstein gravity equation without electromagnetic contributions which are neglected due to scale separation. At this level of approximation, the electromagnetic field is a slave of gravity, namely it does not affect gravity yet affected by it. Indeed, Eq.~(\ref{eq:Mfinal}) is a modified Maxwell equation that includes an explicit interaction term with gravity and thus breaks gauge invariance of the potential. However, Eq.~(\ref{eq:scale-sep2}) shows that this additional source-like term is negligible compared to the current density. Moreover, Eq.~(\ref{eq:Efinal}) shows that $R_{\mu\nu}=0$ out of matter allowing to recover the usual Maxwell equations and to restore the gauge invariance in these regions. Therefore, the main effect of this interaction term on the potential $A_\mu$ is confined to regions filled with matter. It is then important at scales where quantum effects dominate. To this purpose, we explore dimensional reduction of the Dirac spinor field within the same approach.

\section{Dimensional reduction of the Dirac spinor field}

In this section, we account for Fermions and apply the same procedure of dimensional reduction to the action involving a spinor field. In 5D curved spacetime, the Dirac action $\hat S_\mathrm{D}$ associated with a kinetic-like energy density of the source field is defined by~\cite{Salam1982,Macias1991,Pollock2010}
\begin{equation}
{\hat S_\mathrm{D}} = -\frac{i\hbar}{2\ell}\int\limits_{ - {\ell \mathord{\left/
 {\vphantom {\ell 2}} \right.
 \kern-\nulldelimiterspace} 2}}^{{\ell \mathord{\left/
 {\vphantom {h 2}} \right.
 \kern-\nulldelimiterspace} 2}} {\mathrm{d}\zeta\int {\mathrm{d}V {\hat{\mathfrak{L}}_\mathrm{D}}} } \;,
 \label{eq:actionD}
 \end{equation}
where $\hat{\mathfrak{L}}_\mathrm{D}$ is a Dirac Lagrangian density given by
\begin{equation}
{\hat{\mathfrak{L}}_\mathrm{D}} =  \bar \Psi {\hat \gamma ^\mathrm{A}}\hat e_{\left( \mathrm{A} \right)}^M{{\hat D}_M}\Psi  -\left(\hat e_{\left( \mathrm{A} \right)}^M{{\hat D}_M}\bar \Psi\right) {\hat \gamma ^\mathrm{A}}\Psi\;.
\label{eq:13}
\end{equation}
Here, $\Psi$ is a 4D spinor field ($\Psi$ is a four-components field both in 5D and 4D), $\hat{e}_{\left( \mathrm{A} \right)}^{M}$ are the vielbeins of the tetrad representation of the metric $\hat g$ (sometimes referred to as the f\"unfbeins \cite{Zee2013}), $\hat{\gamma}^\mathrm{A}$ are the 5D Dirac matrices and $\hat D _M$ is the 5D covariant derivative for Fermionic fields. The detailed definitions of these quantities are given in Appendix~\ref{Perturb_Dirac}. Finally, notice that the coefficient $1/\ell$ in front of the integral of Eq.~(\ref{eq:actionD}) stems from the same dimensional arguments as for the 5D action of the gravitational field $\hat S_\mathrm{G}$.

To perform the expansion of ${\hat{\mathfrak{L}}_\mathrm{D}}$ in powers of $\zeta$, we assume that the relevant metric $\hat g_{MN}$ from which the vielbeins $\hat{e}_{\left( \mathrm{A} \right)}^{M}$ are derived is given by Eq.~(\ref{eq:2}). The lengthy computations are deferred to Appendix~\ref{Perturb_Dirac} and we just show the final result for ${\hat{\mathfrak{L}}_\mathrm{D}}$ which is given by
\begin{equation}
{\hat{\mathfrak{L}}_D} =  {\bar \Psi \left( {{\gamma ^a}e_{\left( a \right)}^\mu {D_\mu } + 2{\gamma ^5}{\partial _4} - b\zeta{\gamma ^5}{A^\mu }{D_\mu }} \right)\Psi  - \mathrm{h.c.}}\;,
\label{eq:24}
\end{equation}
where $\mathrm{h.c.}$ denotes the Hermitian conjugate. At this stage, one should make assumptions regarding the spinor fields. 
A plausible assumption would be that $\Psi$ does not depend on the fifth dimension, namely $\Psi \left( {x^A} \right)={{\Psi }}\left( {{x}^{\mu }} \right)$ and thus ${{\partial }_{4}}\Psi={{\partial }_{4}}\bar \Psi =0$. In this case, the second term in Eq.~(\ref{eq:24}) drops out. The third term also cancels upon integration over $\zeta$. Therefore, one ends up with
\begin{equation}
{\hat S_D} \equiv S_D= -\frac{i\hbar}{2}\int {\mathrm{d}V\left[ {{{\bar \Psi }}{\gamma ^\mathrm{a}}e_{\left( \mathrm{a} \right)}^\mu {D_\mu }{\Psi} - \left(e_{\left( \mathrm{a} \right)}^\mu {D_\mu }{{\bar \Psi }}\right){\gamma ^\mathrm{a}}{\Psi}} \right]} \;,
\label{eq:29}
\end{equation}
namely the Dirac action $\hat S_D$ becomes identical to the 4D case $S_D$ without further contributions involving the electromagnetic field $A_\mu$.

The coupling of the spinor field with $A_\mu$ and the mass term in Dirac equation can be retrieved by replacing in Eq.~(\ref{eq:12ter}) mass and current densities with their quantum counterparts. They are readily given by
\begin{equation}
\rho_m =m \,\bar \Psi \Psi\;,\quad J^\mu =ec\,\bar \Psi \gamma^\mu\Psi\;,
\end{equation}
and the quantisation of the action $S_m$ reads
\begin{equation}
{{S}_{m}} = \int\limits_{}^{} \mathrm{d}V\sqrt { - g} \left(m c \,\bar \Psi \Psi - \frac{e}{c} \bar \Psi \gamma^\mu A_\mu\Psi \right)\;.
\label{eq:12_dirac}
\end{equation}
Therefore, we end up with all terms involving Dirac spinor fields without invoking additional assumptions. These results strengthen the dimensional reduction approach inspired by elasticity theory of thin plates.

\section{Conclusion}

In this paper, we identify a scale separation between the gravitational and electromagnetic Lagrangian densities of EM action which is similar to that between the stretching and bending energy densities involved in describing the elastic response of thin plates. From this observation, we argue that the KK compactification formalism belongs to the class of techniques as the dimensional reduction of 3D elasticity leading to FvK equations. We exploit this analogy to develop a KK formalism in the framework of elasticity theory of thin plates in which the gravitational and electromagnetic fields are associated with stretching-like and bending-like deformations respectively.
This starting point is different from the classical KK theory where the 5D metric is independent of the fifth component $\zeta$ and the cylindrical or compactification hypothesis is required. We show that the same procedure of dimensional reduction allows us to retrieve the Lagrangian densities of the different fields (gravitational, electromagnetic and Dirac spinors) and of the source terms of matter content (gravitational and electromagnetic). When one solves the equations using a proposed perturbation scheme under the assumption of scale separation, one recovers the Einstein gravitational equations without the electromagnetic energy momentum tensor and the current parts and Maxwell equations with an additional explicit interaction term with gravity.

Our approach suggests that fields and matter are different physical entities, the former being a manifestation of the latter, and both should independently undergo dimensional reduction. Concerning the nature of matter, our scheme suggests that mass and charge in 4D spacetime are two distinct manifestations of the same 5D matter content. While the parameter $\beta\ell$ and 5D matter distribution along the fifth dimension introduced in Sec.~\ref{sec:matter} cannot be fully determined within our approach, the interpretation of our results remain robust as long as $\beta\ell\ll 1$. Moreover, the present study is not dependent on the recent suggestion in~\cite{MAB} that $\ell$ could be as large as a cosmic length scale.

The dimensional reduction of 5D gravitational field induces an explicit term $\propto R_{\mu\nu}A^\mu A^\nu$ that breaks gauge invariance of the electromagnetic potential. However, such a term violates the strong principle of equivalence but not the weak version of it~\cite{Li2016}. Effectively, $R_{\mu\nu}=0$ both in the limit of a Minkowski spacetime and in regions devoid of matter, which restores the gauge invariance for situations where it has been unequivocally established. Our approach produces an explicit interaction between gravity and electromagnetic fields that is embedded in a single term (e.g., by construction a term proportional to $ R A^2$ is discarded) for which both sign and amplitude are uniquely determined.

From the observation that the new interaction term comes into play within the matter content, we extend our approach to Dirac spinor fields. We find that the dimensional reduction of the corresponding 5D Lagrangian density retrieves the 4D one without any additional terms such as anomalous interactions discussed by Pauli \cite{Pollock2010,Salam1982}. The terms involving mass and coupling with the electromagnetic field are simply recovered from the action of the matter by substituting the matter and current densities with their quantum equivalent. 

Finally, the question of a possible unified description of fundamental interactions within general relativity is often intertwined with the dimensionality of our physical world. In this context, Ref.~\cite{MAB} postulated that the current four dimensionality of our physical world results from a dynamical evolution of the spacetime dimension of a matter dominated expanding universe. This result gives a rationale to KK-like approaches: fields mediating other types of interactions should be manifestations of hidden dimensions. Our scheme is then consistent with the scenario that a 4D spacetime is the stable dimension of our universe~\cite{MAB} and that the higher dimensions should appear as perturbations around this fundamental structure. Indeed, one could generalise our approach to include perturbations of any number of passive dimensions paving the way for including more than the electromagnetic field into higher dimensional gravity. The relevance of our approach should be confronted with the results of such a generalization.

\begin{acknowledgments}

This work was supported by the International Research Project ``Non-Equilibrium Physics of Complex Systems'' (IRP-PhyComSys, France-Israel).

%Additional bibliography~\cite{Kramer1997,Chang1976,DiBonna2001,Forgera2004,Francesco2005,Klein1946,Sotiriou2010,Venkataramani2000,Wald,Weinberg}

\end{acknowledgments}

%\bibliographystyle{apsrev4-1}
%\bibliographystyle{unsrt}
%\bibliography{Kaluza-Klein.bib}

\begin{appendix}

\section{On the elasticity theory of thin plates}
\label{app:FvK}

This section briefly outlines the dimensional reduction of 3D bulk elasticity to the case of thin plates. A detailed presentation of this subject can be found in~\cite{LL-Elasticity,Kramer1997,Audoly,Witten2007}.

Consider a 3D homogeneous isotropic solid body regarded as a continuous medium. Under the action of applied forces, the medium undergoes elastic deformations described by an embedding $\vec{r}(x)=[x_i+{u}_i(x)]$, where $x_i$ ($i=1,2,3$) are Euclidean material points of the rest state and $\vec{u}$ is the associated displacement vector field. Upon deformation, the distance between nearby points is given by $\mathrm{d}s^2= g_{ij}\mathrm{d}x_i \mathrm{d}x_j$, where
\begin{equation}
g_{ij}=\frac{d \vec{r}}{d x_i}\cdot \frac{d\vec{r}}{dx_j}=\delta_{ij}+2 u_{ij}\;,
\end{equation}
and $u_{ij}$ is the (dimensionless) symmetric strain tensor. Upon deformations, the increase in elastic energy is assumed to depend only on the distances between nearby pairs of points. In this case, the most general energy functional ${\cal E}[\vec{r}]$ quadratic in the strains and consistent with a homogeneous isotropic material is
\begin{equation}
{\cal E}[\vec{r}]=\frac{1}{2}\int \mathrm{d}x^3 \left[\lambda \left(u_{ii}\right)^2+2\mu  u_{ij}u_{ij}\right]\;,
\label{eq:energy_bulk}
\end{equation}
where the elastic properties of the material are encoded in the Lam\'e coefficients $\lambda$ and $\mu$.

A thin plate is characterized by a very small extent (thickness) $t$ along one dimension of the solid compared to the two other dimensions. The 3D geometry of the plate is described using a 2D manifold associated with its center surface, or centroid, using the embedding $\vec{r}_c(x)= [x_\alpha+u_\alpha(x),w(x)]$ ($\alpha=1,2$) which distinguishes the displacements within the surface ($u_\alpha$) from the one normal to it ($\vec{r}_c\cdot \vec{n}$). In the weakly strained regime, the curvature tensor of the centroid is defined by $C_{\alpha\beta}\equiv \vec{n}\cdot (\partial^2\vec{r}_c/\partial x_\alpha\partial x_\beta)$ ($\alpha,\beta=1,2$). Dimensional reduction of the elastic energy functional for the centroid is found by integrating out the components of the strain tensors which are transverse to the long directions. The mathematical analysis of these approximations and their range of validity is the subject of the theory of elastic shells.

To derive the elastic energy functional for the plate, one starts from Taylor expansion of the embedding $\vec{r}(x)$ in powers of $x_3$. After integration of Eq.~(\ref{eq:energy_bulk}) across the thickness, the energy of such a sheet may be expressed in terms of the in-surface strain tensor $u_{\alpha\beta}$ and the curvature tensor $C_{\alpha\beta}$ of the centroid \cite{LL-Elasticity}. To lowest order in these tensors, ${\cal E}[\vec{r}]$ takes the form
\begin{equation}
{\cal E}[\vec{r}]=S[\mathbf{u}]+B[\mathbf{C}]=\frac{1}{2}\int \mathrm{d}x^2\left[\hat{\lambda} \left(u_{\alpha\alpha}\right)^2+2 \hat{\mu} u_{\alpha\beta}u_{\alpha\beta}\right]+\frac{\kappa}{2} \int   \mathrm{d}x^2\,C_{\alpha\beta}C_{\alpha\beta}\;,
\label{eq:energy_plate}
\end{equation}
where
\begin{equation}
\hat{\lambda}=\frac{E\nu t}{1-\nu^2};\quad \hat{\mu}=\frac{E t}{2(1+\nu)};\quad\kappa=\frac{E t^3}{12(1-\nu^2)}\;.
\end{equation}
The Young modulus $E$ and Poisson ratio $\nu$ are material constants related to Lam\'e coefficients \cite{LL-Elasticity}. Eq.~(\ref{eq:energy_plate}) shows that the energy is the sum of a stretching energy $S[\mathbf{u}]$ involving only the in-surface strain and a bending energy $B[\mathbf{C}]$ involving only the curvature. Notice that Eq.~(\ref{eq:energy_plate})  does not explicitly couple the strains to the curvatures of the manifold and that the scale separation between $B$ and $S$ is mediated by the length scale $\sqrt{\kappa/(E t)}\sim t$.

The minimization of the elastic energy functional can be performed in two ways. The first one uses in-plane and out-of-plane displacements as independent variables. In this case, functional derivatives are taken with respect to $u_\alpha$ and $w$~\cite{LL-Elasticity}. The second route, which is more similar to the minimization of EM action, instead uses the field variables of in-plane strains $u_{\alpha\beta}$ and intrinsic curvature $C_{\alpha\beta}$~\cite{Kramer1997}. However, the strains and the curvatures are implicitly coupled because they are both defined via derivatives of the embedding of the centroid. Therefore, contrary to minimization of EM action, one has to include the constraint imposed by Gauss's Theorema Egregium \cite{Kramer1997}. Of course, both schemes yield the correct FvK equations which can be written in the form
\begin{eqnarray}
\kappa \Delta^2 w &=&2t\, [w,\chi]\;,\\
\Delta^2\chi &=&-E\,[w,w]\;,
\end{eqnarray}
where $\Delta$ is the 2D Laplacian, the brackets $[,]$ are defined by
\begin{equation}
[a,b]=\frac{1}{2}a_{,x_1x_1}b_{,x_2x_2}+\frac{1}{2}a_{,x_2x_2}b_{,x_1x_1}-a_{,x_1x_2}b_{,x_1x_2}\;,
\end{equation} 
and $\chi$ is the so-called Airy stress function \cite{Kramer1997,LL-Elasticity}. 

In many cases, the scale separation between bending and stretching leads the plate to deform in such a way as to bend at large scales and to localize stretching at singular points or along ridges leading to the phenomenon of stress focusing in elastic sheets~\cite{Witten2007} for which crumpled paper is the archetype.

\section{Determination of the appropriate 5D metric tensor $\hat g$}
\label{Perturb_KK}

We are interested in the expansion of the Ricci scalar curvature $\hat R$ in powers of $\zeta\equiv x^4$ up to order ${{\zeta}^{2}}$. To this  purpose, one needs to consider the perturbation of the metric tensor $\hat{g}_{MN}$ up to order ${{\zeta}^{4}}$ as included in Eq.~(\ref{eq:1}). The contravariant metric tensor is obtained by using the identity $\hat{g}^{MP}\hat{g}_{PN}=\hat{\delta}^M_N+ {\cal O}(\zeta^5)$. One shows that $\hat{g}^{MN}$ is given by 
\begin{equation}
{\hat{g} ^{MN}} = \left( {\begin{array}{*{20}{c}}
  {{g^{\mu \nu }} - \left( {{b^2} + a} \right){\zeta^2}{A^\mu }{A^\nu } + \bar d{\zeta^4}\left( {{A^2}} \right){A^\mu }{A^\nu }}&{b\zeta{A^\mu } + \bar e{\zeta^3}\left( {{A^2}} \right){A^\mu }} \\ 
  {b\zeta{A^\mu } + \bar e{\zeta^3}\left( {{A^2}} \right){A^\mu }}&{ - 1 + \left( {{b^2} - c} \right){\zeta^2}\left( {{A^2}} \right) + \bar f{\zeta^4}{{\left( {{A^2}} \right)}^2}} 
\end{array}} \right) +{\cal O}(\zeta^5)\;,
\label{eq:31}
\end{equation}
with
\begin{eqnarray}
\bar d &=&  - \bar a - 2b\bar b + {b^2}\left( {{b^2} + 2a - c} \right) + {a^2}\;, \hfill \\
\bar e &=& b\left( { - {b^2} - a + c} \right) + \bar b\;, \hfill \\
\bar f &=&  - \bar c - 2b\bar b + {b^2}\left( {{b^2} + a - 2c} \right) + {c^2}\;. \hfill  
\label{eq:32}
\end{eqnarray}

Now, we calculate the determinant of the metric tensor given by Eq.~(\ref{eq:1}). Notice that we are interested in the expansion of the determinant up to second order in $\zeta$. We can achieve this by expanding the full metric tensor as $\hat{g} ={{\hat{g} }_{0}}+\zeta{{\hat{g} }_{1}}+{{\zeta}^{2}}{{\hat{g} }_{2}}$, with 
\begin{equation}
{\hat{g} _0} = \left( {\begin{array}{*{20}{c}}
  {{g_{\mu \nu }}}&0 \\ 
  0&{ - 1} 
\end{array}} \right)\;,\qquad {\hat{g} _1} = \left( {\begin{array}{*{20}{c}}
  0&{b{A_\mu }} \\ 
  {b{A_\mu }}&0 
\end{array}} \right)\;,\qquad {\hat{g} _2} = \left( {\begin{array}{*{20}{c}}
  {a{A_\mu }{A_\nu }}&0 \\ 
  0&{c{A^2 }} 
\end{array}} \right)\;.
\label{eq:34}
\end{equation}
Using the identity $\det \hat{g} ={{e}^{\mathrm{Tr} \ln \hat{g} }}$, which is true for any non-singular matrix, one finds
\begin{equation}
\det\hat{g} \approx \det {\hat{g} _0}\left\{ {1 + \mathrm{Tr}\left( {\hat{g} _0^{ - 1}{\hat{g} _1}} \right)\zeta + \left[ {\mathrm{Tr}\left( {\hat{g} _0^{ - 1}{\hat{g} _2}} \right) - \frac{1}{2} {\mathrm{Tr}\left( {{{\left( {\hat{g} _0^{ - 1}{\hat{g} _1}} \right)}^2}} \right) +\frac{1}{2} {{\left( {\mathrm{Tr}\left( {\hat{g} _0^{ - 1}{\hat{g} _1}} \right)} \right)}^2}} } \right]{\zeta^2} } \right\}+ {\cal O}\left( {{\zeta^3}} \right)\;,
\label{eq:37}
\end{equation}
which yields
\begin{equation}
\det \hat{g}  = -\left(\det {{g}}\right)\left[ {1 + \left( {a - c + {b^2}} \right){\zeta^2}A^2} \right] + {\cal O}\left( {{\zeta^3}} \right)\;.
\label{eq:38}
\end{equation}
Therefore, the leading order correction to the determinant of the 4D metric tensor is quadratic in $\zeta$. Using the notation $ g \equiv \det g $, one gets
\begin{equation}
\sqrt{\left|\hat{g}\right|}   = \sqrt { - g} \left[ {1 + \frac{1}{2}\left( {a - c + {b^2}} \right){\zeta^2}A^2} \right] + {\cal O}\left( {{\zeta^3}} \right)\;.
\label{eq:39}
\end{equation}

Let us now work on the Christoffel symbols given by
\begin{equation}
\hat \Gamma _{MN}^R = \frac{1}{2}{\hat{g} ^{RP}}\left( {{\partial _M}{\hat{g} _{PN}} + {\partial _N}{\hat{g} _{PM}} - {\partial _P}{\hat{g} _{MN}}} \right)\;,
\label{eq:4}
\end{equation}
and expand each component of $\hat\Gamma$ up to the appropriate power of $\zeta$ such that all the contributions to the Ricci scalar curvature up to $\zeta^2$ are taken into account. After lengthy algebraic calculations, one finds
\begin{eqnarray}
\hat \Gamma _{\mu \nu }^\rho  &=& \Gamma _{\mu \nu }^\rho  + \frac{1}{2}a{\zeta^2}{g^{\rho \sigma }}\left( {{A_\mu }{F_{\nu \sigma }} + {A_\nu }{F_{\mu \sigma }}} \right) \nonumber \\
&+& \frac{1}{2}\left( {{b^2} + a} \right){\zeta^2}{A^\rho }\left( {{A_{\mu ;\nu }} + {A_{\nu ;\mu }}} \right) - ab{\zeta^2}{A^\rho }{A_\mu }{A_\nu } +{\cal O}(\zeta^3) \;, \label{eq:40}\\
\hat \Gamma _{\mu 4}^4 &=& \Gamma _{4\mu }^4 = \frac{1}{2}{b^2}{\zeta^2}{A^\nu }{F_{\mu \nu }} + ab{\zeta^2}{A_\mu }\left( {{A^2}} \right) - \frac{1}{2}c{\zeta^2}{\left( {{A^2}} \right)_{;\mu }} +{\cal O}(\zeta^3)\;, \\
\hat \Gamma _{44}^\mu  &=& b{A^\mu } + b\left( { - {b^2} - a + c} \right){\zeta^2}{A^\mu }\left( {{A^2}} \right) - \frac{1}{2}c{\zeta^2}{g^{\mu \nu }}{\left( {{A^2}} \right)_{;\nu }}+{\cal O}(\zeta^3) \;,\\
\hat \Gamma_{44}^4 &=& \left( {{b^2} - c} \right)\zeta\left( {{A^2}} \right) +{\cal O}(\zeta^3)\;,\\
\hat \Gamma _{\mu \nu }^4 &=&  - \frac{1}{2}b\zeta\left( {{A_{\nu ;\mu }} + {A_{\mu ;\nu }}} \right) - \frac{1}{2}\left( {\bar b - b\left( {{b^2} + a - c} \right)} \right){\zeta^3}\left( {{A^2}} \right)\left( {{A_{\nu ;\mu }} + {A_{\mu ;\nu }}} \right) \nonumber\\
&+& a\zeta{A_\mu }{A_\nu } + \left[ {2\bar a - a\left( {{b^2} - c} \right)} \right]{\zeta^3}\left( {{A^2}} \right){A_\mu }{A_\nu } \nonumber \\
&-& \frac{1}{2}\bar b{\zeta^3}\left[ {{{\left( {{A^2}} \right)}_{;\mu }}{A_\nu } + {{\left( {{A^2}} \right)}_{;\nu }}{A_\mu }} \right] + \frac{1}{2}ab{\zeta^3}{A^\lambda }\left( {{A_\nu }{F_{\mu \lambda }} + {A_\mu }{F_{\nu \lambda }}} \right) +{\cal O}(\zeta^4)\;, \\
\hat \Gamma _{4\nu }^\mu  &=& a\zeta{A^\mu }{A_\nu } + \left[ {2\bar a + a\left( { - {b^2} - a} \right)} \right]{\zeta^3}\left( {{A^2}} \right){A^\mu }{A_\nu } +\nonumber \\
&+& \frac{1}{2}b\zeta{g^{\mu \lambda }}{F_{\nu \lambda }} + \frac{1}{2}\bar b{\zeta^3}{g^{\mu \lambda }}\left( {{A^2}} \right){F_{\nu \lambda }} + \frac{1}{2}b\left( { - {b^2} - a} \right){\zeta^3}{A^\mu }{A^\lambda }{F_{\nu \lambda }} \nonumber \\
&+& \frac{1}{2}\bar b{\zeta^3}{A^\mu }{\left( {{A^2}} \right)_{;\nu }} - \frac{1}{2}\bar b{\zeta^3}{g^{\mu \lambda }}{\left( {{A^2}} \right)_{;\lambda }}{A_\nu } + \frac{1}{2}bc{\zeta^3}{A^\mu }{\left( {{A^2}} \right)_{;\nu }} +{\cal O}(\zeta^4)\;,
\label{eq:42}
\end{eqnarray}
where the definition of the field tensor $F_{\mu\nu}\equiv A_{\nu;\mu}-A_{\mu;\nu}$ has been used. Notice that Eqs.~(\ref{eq:40}-\ref{eq:42}) do not involve terms proportional to $\bar{c}$ which appear at higher order in $\zeta$.

Now let us turn to the computation of Ricci tensor defined as
\begin{equation}
{\hat R_{MN}} \equiv {\partial _R}\hat \Gamma _{MN}^R - {\partial _N}\hat \Gamma _{MR}^R + \hat \Gamma _{MN}^R\hat \Gamma _{RP}^P - \hat \Gamma _{MP}^R\hat \Gamma _{NR}^P\;.
\label{eq:6}
\end{equation}
Using the expressions for the $\hat\Gamma$'s given by (\ref{eq:40}-\ref{eq:42}), one finds
\begin{eqnarray}
{{\hat R}_{\mu \nu }} &=& {R_{\mu \nu }} + a{A_\mu }{A_\nu } - \frac{1}{2}b\left( {{A_{\nu ;\mu }} + {A_{\mu ;\nu }}} \right) + \nonumber\\
&+& {\left[ {\frac{1}{2}a{\zeta^2}{g^{\rho \sigma }}\left( {{A_\mu }{F_{\nu \sigma }} + {A_\nu }{F_{\mu \sigma }}} \right) + \frac{1}{2}\left( {{b^2} + a} \right){\zeta^2}{A^\rho }\left( {{A_{\mu ;\nu }} + {A_{\nu ;\mu }}} \right)} \right]_{;\rho }} \nonumber \\
&-& ab{\zeta^2}{A^\rho }_{;\rho }{A_\mu }{A_\nu } + \left[ {6\bar a - a\left( {a + 2{b^2} - 2c} \right)} \right]{\zeta^2}\left( {{A^2}} \right){A_\mu }{A_\nu } \nonumber \\
&-& \frac{1}{2}\left( {{b^2} + a - c} \right){\zeta^2}{\left( {{A^2}} \right)_{;\mu ;\nu }} +
b\left( {{b^2} + a - c} \right){\zeta^2}\left( {{A^2}} \right)\left( {{A_{\nu ;\mu }} + {A_{\mu ;\nu }}} \right) \nonumber \\
&+& \frac{3}{2}ab{\zeta^2}{A^\rho }\left( {{A_\mu }{F_{\nu \rho }} + {A_\nu }{F_{\mu \rho }}} \right) 
+ \frac{1}{2}{b^2}{\zeta^2}{g^{\rho \lambda }}\left( {{A_{\nu ;\rho }}{F_{\mu \lambda }} + {F_{\mu \rho }}{F_{\nu \lambda }} + {A_{\mu ;\rho }}{F_{\nu \lambda }}} \right) \nonumber \\
&-& \frac{3}{2}\bar b{\zeta^2}\left\{ {{{\left[ {\left( {{A^2}} \right){A_\nu }} \right]}_{;\mu }} + {{\left[ {\left( {{A^2}} \right){A_\mu }} \right]}_{;\nu }}} \right\} + {\cal O}\left( {{\zeta^3}} \right) \;, \\
{{\hat R}_{4\mu }} &=& a\zeta{A_\mu }{A^\rho }_{;\rho } + \frac{1}{2}\zeta b{\left( {{g^{\rho \lambda }}{F_{\mu \lambda }}} \right)_{;\rho }} + \zeta ab{A_\mu }\left( {{A^2}} \right) - \nonumber\\
&-& \frac{1}{2}\left( {{b^2} + a} \right)\zeta{\left( {{A^2}} \right)_{;\mu }} + \zeta\left( {\frac{1}{2}{b^2} - a} \right){A^\rho }{F_{\mu \rho }} + {\cal O}\left( {{\zeta^3}} \right) \;, \\
{{\hat R}_{44}} &=& b{A^\rho }_{;\rho } + b\left( { - {b^2} - a + c} \right){\zeta^2}\left( {{A^2}} \right){A^\rho }_{;\rho } - \frac{1}{2}c{\zeta^2}{g^{\rho \lambda }}{\left( {{A^2}} \right)_{;\lambda ;\rho }}\nonumber\\
&-& a\left( {{A^2}} \right) + \left[ { - 6\bar a + a\left( {2{b^2} + 2a - c} \right)} \right]{\zeta^2}{\left( {{A^2}} \right)^2} - \frac{1}{4}{b^2}{\zeta^2}{F^{\rho \sigma }}{F_{\rho \sigma }} \nonumber \\
&-& \frac{1}{2}b\left( {{b^2} + a} \right){\zeta^2}{A^\rho }{\left( {{A^2}} \right)_{;\rho }} + {\cal O}\left( {{\zeta^3}} \right) \;. 
\label{eq:45}
\end{eqnarray}

Consequently, the expansion in powers of $\zeta$ of the scalar curvature $\hat R = {\hat{g} ^{MN}}{{\hat R}_{MN}} $ is given by
\begin{eqnarray}
\hat R &=& R - \left( {{b^2} + a} \right){\zeta^2}{A^\mu }{A^\nu }{R_{\mu \nu }} + \frac{3}{4}{b^2}{\zeta^2}{F^{\mu \nu }}{F_{\mu \nu }} \nonumber\\
&+& 2a\left( {{A^2}} \right) + 4\left[ {3\bar a - a\left( {2{b^2} + a - c} \right)} \right]{\zeta^2}{\left( {{A^2}} \right)^2} \nonumber\\
&+& 4b\left( b^2 + a - c \right) \zeta^2 \left( {{A^2}} \right)\left( {{A^\mu }_{;\mu }} \right) 
- {\left[ {2b{A^\rho } + 3\bar b{\zeta^2}\left( {{A^2}} \right){A^\rho }} \right]_{;\rho }}\nonumber\\
&+& \left( {{b^2} + a} \right){\zeta^2}{\left( {{A_\mu }{F^{\mu \rho }} + {A^\rho }{A^\mu }_{;\mu }} \right)_{;\rho }} 
- \frac{1}{2}\left( {{b^2} + a - 2c} \right){\zeta^2}g^{\mu\rho}  {\left( {{A^2}} \right)}_{;\mu;\rho} + {\cal O}\left( {{\zeta^3}} \right) \;.
\label{eq:46}
\end{eqnarray}
Eq.~(\ref{eq:46}) shows that the leading order correction to the 4D scalar curvature is quadratic in $\zeta$.

Finally, we are interested in the the expansion of $\sqrt{\left|\hat{g}\right|}\hat R$ up to order $\zeta^2$. Using Eq.~(\ref{eq:39}) and Eq.~(\ref{eq:46}), one finds that
\begin{eqnarray}
\sqrt{\left|\hat{g}\right|} \,\hat R &=& \sqrt{-g} R\left[ {1 + \frac{1}{2}\left( {{b^2} + a - c} \right)\left( {{A^2}} \right){\zeta^2}} \right] - \left( {{b^2} + a} \right){\zeta^2}{A^\mu }{A^\nu }{R_{\mu \nu }} + \frac{3}{4}{b^2}{\zeta^2}{F^{\mu \nu }}{F_{\mu \nu }} \nonumber \\
&+& 2a\left( {{A^2}} \right) + \left[ {12\bar a - a\left( {7{b^2} + 3a - 3c} \right)} \right]{\zeta^2}{\left( {{A^2}} \right)^2} \nonumber \\
&+& 3b\left( b^2 + a - c \right) \zeta^2\left( {{A^2}} \right)\left( {{A^\mu }_{;\mu }} \right)
-{\left[ {2b{A^\rho } + 3\bar b{\zeta^2}\left( {{A^2}} \right){A^\rho }} \right]_{;\rho }} \nonumber \\
&+& \left( {{b^2} + a} \right){\zeta^2}{\left( {{A_\mu }{F^{\mu \rho }} + {A^\rho }{A^\mu }_{;\mu }} \right)_{;\rho }} - \frac{1}{2}\left( {{b^2} + a - 2c} \right){\zeta^2}g^{\mu\rho}  {\left( {{A^2}} \right)}_{;\mu;\rho} + {\cal O}\left( {{\zeta^3}} \right)\;.  
\label{eq:47}
\end{eqnarray}

Now, let us consider the action given by Eq.~(\ref{eq:action1}). Using Gauss theorem, terms in Eq.~(\ref{eq:47}) that involve total 4D derivatives can be transformed into an integral over the hypersurface surrounding the whole 4D volume. Assuming that these terms do not contribute to the action, they can thus be discarded from Eq.~(\ref{eq:47}) yielding
\begin{eqnarray}
\sqrt{\left|\hat{g}\right|} \,\hat R &=& \sqrt{-g} R\left[ {1 + \frac{1}{2}\left( {{b^2} + a - c} \right)\left( {{A^2}} \right){\zeta^2}} \right]
- \left( {{b^2} + a} \right){\zeta^2}{A^\mu }{A^\nu }{R_{\mu \nu }} + \frac{3}{4}{b^2}{\zeta^2}{F^{\mu \nu }}{F_{\mu \nu }} \nonumber \\
&+& 2a\left( {{A^2}} \right) + \left[ {12\bar a - a\left( {7{b^2} + 3a - 3c} \right)} \right]{\zeta^2}{\left( {{A^2}} \right)^2} \nonumber \\
&+& 3b\left( b^2 + a - c \right) \zeta^2 \left( {{A^2}} \right)\left( {{A^\mu }_{;\mu }} \right) + {\cal O}\left( {{\zeta^3}} \right) \;.
\label{eq:48}
\end{eqnarray}
Notice that Eq.~(\ref{eq:48}) is independent of $\bar b$. We aim at identifying the vector field $A_\mu$ with the electromagnetic potential field. To this purpose, we should build an action that allows for $A_\mu$ to satisfy differential equations linear in $A_\mu$. Therefore, under the integral sign for the action there must stand an expression quadratic in that field. Moreover, we impose that the potentials enter into the expression of the action $\hat{S}_{\mathrm{G}}$ only through their derivatives $A_{\mu;\nu}$~\cite{LL1980}. To fulfil these conditions, one should cancel the undesirable terms in Eq.~(\ref{eq:48}) by fixing some of the constants in the expansion of the 5D metric tensor given by Eq.~(\ref{eq:1}). These physical constraints allow us to impose
\begin{equation}
a=\bar{a}=0\;;\qquad   c=b^2\;.
\end{equation}
Interestingly the condition $a=0$ implies that the field $A^\mu$ is massless. Using these conditions, Eq.~(\ref{eq:48}) is simplified into
\begin{equation}
\frac{\sqrt{\left|\hat{g}\right|}}{\sqrt{-g}} \,\hat R = R  + \frac{3}{4}{b^2}{\zeta^2}{F^{\mu \nu }}{F_{\mu \nu }} - {b^2}{\zeta^2}{A^\mu }{A^\nu }{R_{\mu \nu }}+ {\cal O}\left( {{\zeta^3}} \right) \;.
\label{eq:49}
\end{equation}
Eq.~(\ref{eq:49}) is the main result of this Section and is reproduced in Eq.~(\ref{eq:10}) in the main text. Notice that using the identity ${A^\mu }R_{\mu \nu }={A^\mu}_{;\nu;\mu}-{A^\mu}_{;\mu;\nu}$, one has 
\begin{equation}
A^\mu {A^\nu }R_{\mu \nu }= \left({A^\mu}_{;\mu}\right)^2-{A^\mu}_{;\nu}{A^\nu}_{;\mu}+\left(A^\nu {A^\mu}_{;\nu}-A^\mu {A^\nu}_{;\nu}\right)_{;\mu}\;.
\label{eq:A27}
\end{equation}
showing explicitly that the third term in the right hand side of Eq.~(\ref{eq:49}) can indeed be rewritten as a quadratic function of the potential derivatives only (up to total derivatives that do not contribute to the action).

The main result of these calculations is that the 5D metric that allows for an expansion of the scalar curvature up to $\zeta^2$, with suitable properties, involves a single unknown constant $b$. We will then focus on the 5D metric tensor given by Eq.~(\ref{eq:2}). Now, we summarize our results (up to the desired order in $\zeta$) using the simple expression for $\hat g _{MN}$. First, one has
\begin{equation}
{\hat{g} ^{MN}} = \left( {\begin{array}{*{20}{c}}
{{g^{\mu \nu }} - {b^2}{\zeta^2}{A^\mu }{A^\nu }}&{b\zeta{A^\mu }}\\
{b\zeta{A^\mu }}&{ - 1}
\end{array}} \right)\;.
\label{eq:3bis}
\end{equation}
Notice that $\hat{g} ^{MN}$ has the same structure as the KK covariant metric tensor~\cite{Overduin1997}. It can be verified that for this metric, the determinant $\left|\hat{g}\right|$ satisfies Eq.~(\ref{eq:10bis}) and the Christoffel symbols become
\begin{eqnarray}
\hat \Gamma _{\mu \nu}^\rho &=& \Gamma _{\mu \nu }^\rho  + \frac{1}{2}{b^2}{\zeta^2}{A^\rho }\left( {{A_{\nu ;\mu }} + {A_{\mu ;\nu }}} \right) \;,
\label{eq:4bis}\\
\hat \Gamma _{\mu 4}^\nu    &=& \hat \Gamma _{4\mu }^\nu  = \frac{1}{2}b\zeta{g^{\nu \rho }}{F_{\mu \rho }} - \frac{1}{2}{b^3}{\zeta^3}{A^\rho }{A^\nu }{F_{\mu \rho }} + \frac{1}{2}{b^3}{\zeta^3}{A^\nu }{\left( {{A^2}} \right)_{;\mu }} \;,\\
\hat \Gamma _{\mu \nu }^4   &=&  - \frac{1}{2}b\zeta\left( {{A_{\nu ;\mu }} + {A_{\mu ;\nu }}} \right) \;,\\
\hat \Gamma _{4\mu }^4      &=& \hat \Gamma _{\mu 4}^4 = \frac{1}{2}{b^2}{\zeta^2}{A^\sigma }{F_{\mu \sigma }} - \frac{1}{2}{b^2}{\zeta^2}{\left( {{A^2}} \right)_{,\mu }} \;,\\
\hat \Gamma _{44}^\mu       &=& b{A^\mu } - \frac{1}{2}{b^2}{\zeta^2}{g^{\mu \sigma }}{\left( {{A^2}} \right)_{,\sigma }} \;,\\
\hat \Gamma _{44}^4 &=& 0\;.
\label{eq:5bis}
\end{eqnarray}
A corollary of the result for $\hat \Gamma _{\mu 4}^\nu $ is that $\hat \Gamma _{4\rho }^\rho  = \frac{1}{2}{b^3}{\zeta^3}{A^\rho }{\left( {{A^2}} \right)_{;\rho }}$. The Ricci tensor is simplified into 
\begin{eqnarray}
{{\hat R}_{\mu \nu }} &=& {R_{\mu \nu }} - \frac{1}{2}b\left( {{A_{\nu ;\mu }} + {A_{\mu ;\nu }}} \right) + {\left[ {\frac{1}{2}{b^2}{\zeta^2}{A^\rho }\left( {{A_{\nu ;\mu }} + {A_{\mu ;\nu }}} \right)} \right]_{;\rho }} \nonumber \\
&& + \frac{1}{4}{b^2}{\zeta^2}{g^{\rho \sigma }}\left[ {{F_{\mu \sigma }}\left( {{A_{\nu ;\rho }} + {A_{\rho ;\nu }}} \right) + {F_{\nu \rho }}\left( {{A_{\sigma ;\mu }} + {A_{\mu ;\sigma }}} \right)} \right]\;,
\label{eq:8bis}\\
{\hat R_{4\mu }} &=& \frac{1}{2}b\zeta{\left( {{g^{\rho \sigma }}{F_{\mu \sigma }}} \right)_{;\rho }} - \frac{1}{2}{b^2}\zeta{\left( {{A^2}} \right)_{;\mu }} + \frac{1}{2}{b^2}\zeta{A^\sigma }{F_{\mu \sigma }}\;,
\label{eq:7bis}\\
{\hat R_{44}} &=& b{A^\rho }_{;\rho } - \frac{1}{2}{b^2}{\zeta^2}{g^{\rho \sigma }}{\left( {{A^2}} \right)_{;\sigma ;\rho }} - \frac{1}{2}{b^3}{\zeta^2}{A^\sigma }{\left( {{A^2}} \right)_{;\sigma }} - \frac{1}{4}{b^2}{\zeta^2}{F^{\mu \nu }}{F_{\mu \nu }}\;.
\label{eq:9bis}
\end{eqnarray}
Finally, the Ricci scalar curvature coincides with Eq.~(\ref{eq:10}), up to total derivatives.

\section{Expansion of the Lagrangian density of Dirac spinor field}
\label{Perturb_Dirac}

This appendix is devoted to the expansion in powers of $\zeta$ of the Dirac Lagrangian density given by Eq.~(\ref{eq:13}). Here, we restrict the computations to the metric given by Eq.~(\ref{eq:2}). Before proceeding, let us define the different quantities introduced in Eq.~(\ref{eq:13}). First,
$\hat{e}_{\left( \mathrm{A} \right)}^{M}$ are the vielbeins which are determined from the tetrad representation of the metric ${\hat{g} _{MN}}$ such that
\begin{equation}
{\hat{g} _{MN}} = {\hat \eta _{\mathrm{AB}}}\hat e_M^{\left( \mathrm{A} \right)}\hat e_N^{\left( \mathrm{B} \right)}\;,
\label{eq:51}
\end{equation}
where ${{\hat{\eta }}_{\mathrm{AB}}}=\mathrm{diag}\left( +1,-1,-1,-1,-1 \right)$. Then, $\hat{\gamma}^\mathrm{A}$ are the 5D Dirac matrices given by
\begin{equation}
\hat{\gamma}^0=\gamma^0\;;\quad\hat{\gamma}^1=\gamma^1\;;\quad\hat{\gamma}^2=\gamma^2\;;\quad\hat{\gamma}^3=\gamma^3\;;\quad\hat{\gamma}^4=\gamma^5\equiv i\gamma^0\gamma^1\gamma^2\gamma^3\;.
\end{equation}
While the $ {{\hat{\gamma }}^{\mathrm{A}}}$ are defined in the tetrad frame, hence they are constant and obey the anti-commutation relations given by
\begin{equation}
\left\{ {{\hat\gamma ^{\mathrm{A}}},{\hat\gamma ^{\mathrm{B}}}} \right\} = 2{\hat \eta ^{\mathrm{AB}}}\;.
\label{eq:14}
\end{equation}
One can also define the curved-space Dirac gamma matrices $\hat \gamma ^M$ as
\begin{equation}
{\hat \gamma^M} = \hat e_{\left( \mathrm{A} \right)}^M{\hat\gamma ^\mathrm{A}} \;,
\label{eq:18}
\end{equation}
\noindent
which obey anti-commutation relations given by
\begin{equation}
\left\{ {{{\hat \gamma }^M},{{\hat \gamma }^N}} \right\} = 2{\hat{g} ^{MN}}.
\label{eq:19}
\end{equation}
Finally, $\hat D _M$ is the 5D covariant derivative for Fermionic fields defined by
\begin{equation}
{\hat D_M} = {\hat{\partial}_M} - \frac{i}{4}\hat \omega _M^{\mathrm{AB}}{\hat{\sigma}_{\mathrm{AB}}}\;,
\label{eq:15}
\end{equation}
where $\hat \omega _M^{\mathrm{AB}}$ are the spin connections  and $\hat{\sigma}_{\mathrm{AB}}$ are the spin operators. They are explicitly given by
\begin{eqnarray}
\hat \omega _M^{\mathrm{AB}} &=& \hat e_N^{(\mathrm{A})}\hat \Gamma _{RM}^N{\hat e^{R\left( \mathrm{B} \right)}} + \hat e_N^{(\mathrm{A})}{\partial _M}{\hat e^{N\left( \mathrm{B} \right)}}\;,
\label{eq:58}\\
{\hat{\sigma}^{\mathrm{AB}}} &=& \frac{i}{2}\left[ {{\hat{\gamma}^\mathrm{A}},{\hat{\gamma}^\mathrm{B}}} \right]\equiv \frac{i}{2}\left( {{\hat{\gamma}^\mathrm{A}}{\hat{\gamma}^\mathrm{B}}}- {{\hat{\gamma}^\mathrm{B}}{\hat{\gamma}^\mathrm{A}}} \right)\;.
\label{eq:16}
\end{eqnarray}
Note that the spin connections are antisymmetric with respect to the exchange of $\mathrm{A}$ and $\mathrm{B}$.

To proceed with the dimensional reduction of the Dirac action for the spinor field, we start by determining the tetrad representation of the metric ${\hat{g} _{MN}}$ given by Eq.~(\ref{eq:2}). We start with a general form of the vielbeins given by
\begin{equation}
\hat e_{\left( \mathrm{A} \right)}^M = \left( {\begin{array}{*{20}{c}}
{e_{\left( \mathrm{a} \right)}^\mu }&{{P^\mu }}\\
{{Q_{\left( \mathrm{a} \right)}}}&{\phi^{-1}}
\end{array}} \right)\;,
\label{eq:52}
\end{equation}
where $P^\mu$ and $Q_{\left( \mathrm{a} \right)}$ are arbitrary 4-vectors, $\phi $ is an arbitrary scalar and $e_{\left( \mathrm{a} \right)}^{\mu }$ are the $4\times 4$ vierbeins that give rise to the 4D metric ${{g}_{\mu \nu }}$. One has
\begin{equation}
\hat e_{\left( \mathrm{A} \right)}^M\hat e_N^{\left( \mathrm{A} \right)} = \delta _N^M\;;\qquad
\hat e_{\left( \mathrm{A} \right)}^M\hat e_M^{\left( \mathrm{B} \right)} = \delta _\mathrm{A}^\mathrm{B}\;,
\label{eq:50}
\end{equation}
with $\hat{e}_{M}^{\left( \mathrm{A} \right)}$ the corresponding $1$-forms of the 5D vielbeins. Using Eq.~(\ref{eq:50}) allows us to write the $1$-forms as
\begin{equation}
\hat e_M^{\left( \mathrm{A} \right)} = \left( {\begin{array}{*{20}{c}}
{e_\mu ^{\left( \mathrm{a} \right)}}&{ - \phi {P^{\left( \mathrm{a} \right)}}}\\
{ - \phi {Q_\mu }}&\phi 
\end{array}} \right)\;.
\label{eq:53}
\end{equation}
The representations of the vielbeins and the $1$-forms given by Eqs.~(\ref{eq:52}),(\ref{eq:53}) yield a metric ${\hat{g} _{MN}}$ that is given by
\begin{equation}
{\hat{g} _{MN}} = \left( {\begin{array}{*{20}{c}}
{{g_{\mu \nu }} - {\phi ^2}{Q_\mu }{Q_\nu }}&{ - \phi {P_\mu } + {\phi ^2}{Q_\mu }}\\
{ - \phi {P_\nu } + {\phi ^2}{Q_\nu }}&{- {\phi ^2}+{\phi ^2}{P^2} }
\end{array}} \right)\;.
\label{eq:54}
\end{equation}
Upon comparison of Eq.~(\ref{eq:54}) with Eq.~(\ref{eq:2}), one concludes that
\begin{equation}
Q_\mu = 0\;;\quad
\phi  = 1\;;\quad
{P_\mu } =  - b\zeta{A_\mu } \;.
\label{eq:55}
\end{equation}
Therefore, the components of the vielbeins and the $1$-forms associated with the metric ${\hat{g} _{MN}}$ are given by
\begin{eqnarray}
&&\hat e_{\left( \mathrm{a} \right)}^\mu  = e_{\left( \mathrm{a} \right)}^\mu \;;\quad\hat e_{\left( 4 \right)}^\mu  =  - b\zeta{A^\mu }\;;\quad\hat e_{\left( \mathrm{a} \right)}^4 = 0\;;\quad\hat e_{\left( 4 \right)}^4 = 1\;,
\label{eq:56a}\\
&&\hat e_\mu ^{\left( \mathrm{a} \right)} = e_\mu ^{\left( \mathrm{a} \right)}\;;\quad\hat e_4^{\left( \mathrm{a} \right)} = b\zeta{A^{\left( \mathrm{a} \right)}}\;;\quad\hat e_\mu ^{\left( 4 \right)} = 0\;;\quad\hat e_4^{\left( 4 \right)} = 1\;.
\label{eq:56b}
\end{eqnarray}
In the following, we will also use the vielbeins with all upper indices, namely ${{\hat{e}}^{M\left( \mathrm{A} \right)}}={{\hat{\eta }}^{\mathrm{AB}}}\hat{e}_{\left( \mathrm{B} \right)}^{M}$. They are given by
\begin{equation}
{\hat e^{\mu \left( \mathrm{a} \right)}} = {e^{\mu \left( \mathrm{a} \right)}}\;;\quad{\hat e^{\mu \left( 4 \right)}} = b\zeta{A^\mu }\;;\quad{\hat e^{4\left( \mathrm{a} \right)}} = 0\;;\quad{\hat e^{4\left( 4 \right)}} =  - 1\;.
\label{eq:57}
\end{equation}

We can now calculate the spin connections $\hat \omega _M^{\mathrm{AB}}$ that are needed in the covariant derivatives of Fermionic fields. It is useful to write down explicitly Eq.~(\ref{eq:58}) to separate the 4D terms from the 5D ones, namely
\begin{eqnarray}
&&\hat \omega _\mu ^{\mathrm{ab}} = \hat e_\nu ^{\left( \mathrm{a} \right)}\hat \Gamma _{\lambda \mu }^\nu {{\hat e}^{\lambda \left( \mathrm{b} \right)}} + \hat e_4^{\left( \mathrm{a} \right)}\hat \Gamma _{\lambda \mu }^4{{\hat e}^{\lambda \left( \mathrm{b} \right)}} + \hat e_\nu ^{\left( \mathrm{a} \right)}\hat \Gamma _{4\mu }^\nu {{\hat e}^{4\left( \mathrm{b} \right)}} + \hat e_4^{\left( \mathrm{a} \right)}\hat \Gamma _{4\mu }^4{{\hat e}^{4\left( \mathrm{b} \right)}} + \hat e_\nu ^{\left( \mathrm{a} \right)}{\partial _\mu }{{\hat e}^{\nu \left( \mathrm{b} \right)}} + \hat e_4^{\left( \mathrm{a} \right)}{\partial _\mu }{{\hat e}^{4\left( \mathrm{b} \right)}}\;,\\
&&\hat \omega _\mu ^{\mathrm{a}4} = \hat e_\nu ^{\left( \mathrm{a} \right)}\hat \Gamma _{\lambda \mu }^\nu {{\hat e}^{\lambda \left( 4 \right)}} + \hat e_4^{\left( \mathrm{a} \right)}\hat \Gamma _{\lambda \mu }^4{{\hat e}^{\lambda (4)}} + \hat e_\nu ^{\left( \mathrm{a} \right)}\hat \Gamma _{4\mu }^\nu {{\hat e}^{4\left( 4 \right)}} + \hat e_4^{\left( \mathrm{a} \right)}\hat \Gamma _{4\mu }^4{{\hat e}^{4\left( 4 \right)}} + \hat e_\nu ^{\left( \mathrm{a} \right)}{\partial _\mu }{{\hat e}^{\nu \left( 4 \right)}} + \hat e_4^{\left( \mathrm{a} \right)}{\partial _\mu }{{\hat e}^{4\left( 4 \right)}}\;,\\
&&\hat \omega _4^{\mathrm{ab}}    = \hat e_\nu ^{\left( \mathrm{a} \right)}\hat \Gamma _{\lambda 4}^\nu {{\hat e}^{\lambda \left( \mathrm{b} \right)}} + \hat e_4^{\left( \mathrm{a} \right)}\hat \Gamma _{\lambda 4}^4{{\hat e}^{\lambda \left( \mathrm{b} \right)}} + \hat e_\nu ^{\left( \mathrm{a} \right)}\hat \Gamma _{44}^\nu {{\hat e}^{4\left( \mathrm{b} \right)}} + \hat e_\nu ^{\left( \mathrm{a} \right)}{\partial _4}{{\hat e}^{\nu \left( \mathrm{b} \right)}} + \hat e_4^{\left( \mathrm{a} \right)}{\partial _4}{{\hat e}^{4\left( \mathrm{b} \right)}}\;,\\
&&\hat \omega _4^{\mathrm{a}4}    = \hat e_\nu ^{\left( \mathrm{a} \right)}\hat \Gamma _{\lambda 4}^\nu {{\hat e}^{\lambda \left( 4 \right)}} + \hat e_4^{\left( \mathrm{a} \right)}\hat \Gamma _{\lambda 4}^4{{\hat e}^{\lambda \left( 4 \right)}} + \hat e_\nu ^{\left( \mathrm{a} \right)}\hat \Gamma _{44}^\nu {{\hat e}^{4\left( 4 \right)}} + \hat e_\nu ^{\left( \mathrm{a} \right)}{\partial _4}{{\hat e}^{\nu \left( 4 \right)}} + \hat e_4^{\left( \mathrm{a} \right)}{\partial _4}{{\hat e}^{4\left( 4 \right)}}\;.
\label{eq:60}
\end{eqnarray}
where ${{\partial }_{4}}$ is the derivative with respect to the fifth dimension $\zeta\equiv x^4$. Using the expressions of the vielbeins given by Eq.~(\ref{eq:57}), one gets
\begin{eqnarray}
\hat \omega _\mu ^{\mathrm{ab}} &=& e_\nu ^{\left( \mathrm{a} \right)}\hat \Gamma _{\lambda \mu }^\nu {e^{\lambda \left( \mathrm{b} \right)}} + b\zeta{A^{\left( \mathrm{a} \right)}}\hat \Gamma _{\lambda \mu }^4{e^{\lambda \left( \mathrm{b} \right)}} + e_\nu ^{\left( \mathrm{a} \right)}{\partial _\mu }{e^{\nu \left( \mathrm{b} \right)}}\;,\\
\hat \omega _\mu ^{\mathrm{a}4} &=& e_\nu ^{\left( \mathrm{a} \right)}\hat \Gamma _{\lambda \mu }^\nu b\zeta{A^\lambda } + b\zeta{A^{\left( \mathrm{a} \right)}}\hat \Gamma _{\lambda \mu }^4b\zeta{A^\lambda } - e_\nu ^{\left( \mathrm{a} \right)}\hat \Gamma _{4\mu }^\nu  - b\zeta{A^{\left( \mathrm{a} \right)}}\hat \Gamma _{4\mu }^4\;,\\
\hat \omega _4^{\mathrm{ab}} &=& e_\nu ^{\left( \mathrm{a} \right)}\hat \Gamma _{\lambda 4}^\nu {e^{\lambda \left( \mathrm{b} \right)}} + b\zeta{A^{\left( \mathrm{a} \right)}}\hat \Gamma _{\lambda 4}^4{e^{\lambda \left( \mathrm{b} \right)}} + e_\nu ^{\left( \mathrm{a} \right)}{\partial _4}{e^{\nu \left( \mathrm{b} \right)}}\;,\\
\hat \omega _4^{\mathrm{a}4} &=& e_\nu ^{\left( \mathrm{a} \right)}\hat \Gamma _{\lambda 4}^\nu b\zeta{A^\lambda } + b\zeta{A^{\left( \mathrm{a} \right)}}\hat \Gamma _{\lambda 4}^4b\zeta{A^\lambda } - e_\nu ^{\left( \mathrm{a} \right)}\hat \Gamma _{44}^\nu  + be_\nu ^{\left( \mathrm{a} \right)}{A^\nu }\left( {{\partial _4}\zeta} \right)\;.
\label{eq:61}
\end{eqnarray}
Using ${{\partial }_{4}}\zeta=1$, $e_{\nu }^{\left( \mathrm{a} \right)}{{\partial }_{4}}{{e}^{\nu \left( \mathrm{b} \right)}}={{\eta }^{\mathrm{ab}}}{{\partial }_{4}}$ (that is the 4D vierbeins do not depend on the fifth coordinate $\zeta$) and $\omega _{\mu }^{\mathrm{ab}}=e_{\nu }^{(\mathrm{a})}\Gamma _{\sigma \mu }^{\nu }{{e}^{\sigma (\mathrm{b})}}+e_{\nu }^{(\mathrm{a})}{{\partial }_{\mu }}{{e}^{\nu (\mathrm{b})}}$, one finds
\begin{eqnarray}
\hat \omega _\mu ^{\mathrm{ab}} &=& \omega _\mu ^{\mathrm{ab}} + b\zeta{A^{\left( \mathrm{a} \right)}}\hat \Gamma _{\lambda \mu }^4{e^{\lambda \left( \mathrm{b} \right)}} + e_\nu ^{(\mathrm{a})}\left( {\hat \Gamma _{\sigma \mu }^\nu  - \Gamma _{\sigma \mu }^\nu } \right){e^{\sigma (\mathrm{b})}}\;,\\
\hat \omega _\mu ^{\mathrm{a}4} &=& b\zeta e_\nu ^{\left( \mathrm{a} \right)}\hat \Gamma _{\lambda \mu }^\nu {A^\lambda } + {b^2}{\zeta^2}{A^{\left( \mathrm{a} \right)}}\hat \Gamma _{\lambda \mu }^4{A^\lambda } - e_\nu ^{\left( \mathrm{a} \right)}\hat \Gamma _{4\mu }^\nu  - b\zeta{A^{\left( \mathrm{a} \right)}}\hat \Gamma _{4\mu }^4\;,\\
\hat \omega _4^{\mathrm{ab}} &=& e_\nu ^{\left( \mathrm{a} \right)}\hat \Gamma _{\lambda 4}^\nu {e^{\lambda \left( \mathrm{b} \right)}} + b\zeta{A^{\left( \mathrm{a} \right)}}\hat \Gamma _{\lambda 4}^4{e^{\lambda \left( \mathrm{b} \right)}} + {\eta ^{\mathrm{ab}}}{\partial _4}\;,\\
\hat \omega _4^{\mathrm{a}4} &=& b\zeta e_\nu ^{\left( \mathrm{a} \right)}\hat \Gamma _{\lambda 4}^\nu {A^\lambda } + {b^2}{\zeta^2}{A^{\left( \mathrm{a} \right)}}\hat \Gamma _{\lambda 4}^4{A^\lambda } - e_\nu ^{\left( \mathrm{a} \right)}\hat \Gamma _{44}^\nu  + b{A^{\left( \mathrm{a} \right)}}\;.
\label{eq:62}
\end{eqnarray}
Now, using the Christoffel symbols as given by Eqs.~(\ref{eq:4bis})-(\ref{eq:5bis}) yields
\begin{eqnarray}
\hat \omega _\mu ^{\mathrm{ab}} &=& \omega _\mu ^{\mathrm{ab}}\;,
\label{eq:63_0}\\
\hat \omega _\mu ^{\mathrm{a}4} &=& \frac{1}{2}b\zeta e_\nu ^{\left( \mathrm{a} \right)}\left( {2\Gamma _{\lambda \mu }^\nu {A^\lambda } + {g^{\nu \lambda }}{F_{\lambda \mu }}} \right)\;,\\
\hat \omega _4^{\mathrm{ab}} &=& {\eta ^{\mathrm{ab}}}{\partial _4} + \frac{1}{2}b\zeta{e^{\rho \left( \mathrm{a} \right)}}{e^{\lambda \left( \mathrm{b} \right)}}{F_{\lambda \rho }}\;,\\
\hat \omega _4^{\mathrm{a}4} &=&  - \frac{1}{2}{b^2}{\zeta^2}{e^{\rho \left( \mathrm{a} \right)}}\left( {{A^\lambda }{F_{\rho \lambda }} - {{\left( {{A^2}} \right)}_{;\rho }}} \right)\;.
\label{eq:63}
\end{eqnarray}

We now turn to the 5D covariant derivative for Fermionic fields defined by Eq.~(\ref{eq:15}). For convenience, let us separate explicitly in Eq.~(\ref{eq:15}) the 4D terms from the 5D one. Using the anti-commutation rules of Dirac matrices, Eq.~(\ref{eq:15}) becomes
\begin{eqnarray}
{{\hat D}_\mu } &=& {\partial _\mu } + \frac{1}{4}\hat \omega _\mu ^{\mathrm{ab}}{\gamma _\mathrm{a}}{\gamma _\mathrm{b}} - \frac{1}{2}\hat \omega _\mu ^{\mathrm{a}4}{\gamma _5}{\gamma _\mathrm{a}}\;,
\label{eq:64a}\\
{{\hat D}_4} &=& {\partial _4} + \frac{1}{4}\hat \omega _4^{\mathrm{ab}}{\gamma _\mathrm{a}}{\gamma _\mathrm{b}} - \frac{1}{2}\hat \omega _4^{\mathrm{a}4}{\gamma _5}{\gamma _\mathrm{a}}\;.
\label{eq:64b}
\end{eqnarray}
Plugging into these equations the spin connections given by Eqs.~(\ref{eq:63_0})-(\ref{eq:63}), the definition of the 4D Fermionic covariant derivative ${{D}_{\mu }}={{\partial }_{\mu }}-\frac{i}{4}\omega _{\mu }^{\mathrm{ab}}{{\sigma }_{\mathrm{ab}}}$ and using the identity ${{\eta }^{\mathrm{ab}}}{{\gamma }_{\mathrm{a}}}{{\gamma }_{\mathrm{b}}}={{\gamma }^{\mathrm{a}}}{{\gamma }_{\mathrm{a}}}=4{{I}_{4}}$, one obtains
\begin{eqnarray}
  {{\hat D}_\mu } &=& {D_\mu } - \frac{1}{4}b\zeta e_\nu ^{\left( \mathrm{a} \right)}\left( {2\Gamma _{\lambda \mu }^\nu {A^\lambda } + {g^{\nu \lambda }}{F_{\lambda \mu }}} \right){\gamma _4}{\gamma _\mathrm{a}} \;,
  \label{eq:22a} \\
  {{\hat D}_4} &=& 2{\partial _4} + \frac{1}{8}b\zeta{e^{\rho \left( \mathrm{a} \right)}}{e^{\lambda \left( \mathrm{b} \right)}}{\gamma _\mathrm{a}}{\gamma _\mathrm{b}}{F_{\lambda \rho }} +  \frac{1}{4}{b^2}{\zeta^2}{e^{\rho \left( \mathrm{a} \right)}}\left[ {{A^\lambda }{F_{\rho \lambda }} - {{\left( {{A^2}} \right)}_{;\rho }}} \right]{\gamma _4}{\gamma _\mathrm{a}}\;.
\label{eq:22b}
\end{eqnarray}
Using Eqs.~(\ref{eq:56a}),(\ref{eq:22a}),(\ref{eq:22b}), the Dirac operator ${{\gamma }^{\mathrm{A}}}\hat{e}_{\left( \mathrm{A} \right)}^{M}{{\hat{D}}_{M}}$ in Eq.~(\ref{eq:13}) becomes
\begin{eqnarray}
{\gamma ^\mathrm{A}}\hat e_{\left( \mathrm{A} \right)}^M{{\hat D}_M} &=& {{\gamma ^\mathrm{a}}e_{\left( \mathrm{a} \right)}^\mu {D_\mu } + 2{\gamma ^5}{\partial _4} - b\zeta{\gamma ^5}{A^\mu }{D_\mu } - } \nonumber\\
 &-& {\frac{1}{8}b\zeta{\gamma ^5}e_\nu ^{\left( \mathrm{b} \right)}{e^{\mu \left( \mathrm{a} \right)}}\left( {4\Gamma _{\lambda \mu }^\nu {A^\lambda } + {g^{\nu \lambda }}{F_{\lambda \mu }}} \right){\gamma_\mathrm{a}}{\gamma_\mathrm{b}} - \frac{1}{2}{b^2}{\zeta^2}{\gamma ^\mathrm{a}}e_{\left( \mathrm{a} \right)}^\rho {A_{\rho ;\lambda }}{A^\lambda }} \;.
\label{eq:23}
\end{eqnarray}
Finally the Dirac Lagrangian density given by Eq.~(\ref{eq:13}) is simplified into Eq.~(\ref{eq:24}). Indeed the two last terms in Eq.~(\ref{eq:23}) do not involve  derivatives of the spinor field with respect to $D_\mu$ or $\partial_4$ and therefore they cancel out.

\end{appendix}

\end{document}